\documentclass[conference]{IEEEtran}
\normalsize
\ifCLASSINFOpdf
\else
\fi

\usepackage{amsthm}
\usepackage{amsmath}
\usepackage{amssymb}
\usepackage{graphicx}
\usepackage{esint}
\usepackage{amsfonts}
\usepackage{cite}
\usepackage{balance}
\usepackage{caption}
\usepackage{epstopdf}
\usepackage{color}
\usepackage{tikz}
\usepackage{graphicx}
\usepackage{fixmath}
\usepackage{pgfplots}
\usepackage{amsmath,amsfonts}
\usepackage{algorithmic}
\usepackage{algorithm}
\usepackage{array}
\usepackage[caption=false,font=normalsize,labelfont=sf,textfont=sf]{subfig}
\usepackage{textcomp}
\usepackage{stfloats}
\usepackage{url}
\usepackage{verbatim}
\pgfplotsset{compat=newest}

\makeatletter

\theoremstyle{plain}
\newtheorem{prop}{\protect\theoremname}

\makeatother

\providecommand{\theoremname}{Proposition}

\theoremstyle{plain}

\makeatother

\providecommand{\lemmaname}{Lemma}

\theoremstyle{plain}
\newtheorem{thm}{\protect\theoremname}

\makeatother

\providecommand{\theoremname}{Theorem}

\DeclareMathOperator{\erf}{erf}

\begin{document}
\title{Modeling blockage in high directional wireless systems }

\author{
\IEEEauthorblockN{Evangelos Koutsonas\IEEEauthorrefmark{1}, Alexandros-Apostolos A. Boulogeorgos\IEEEauthorrefmark{1},  Stylianos E. Trevlakis\IEEEauthorrefmark{2}, \\
Tanweer Ali\IEEEauthorrefmark{3}, and Theodoros A. Tsiftis\IEEEauthorrefmark{4}\IEEEauthorrefmark{5} \\
\IEEEauthorrefmark{1} Department of Electrical and Computer Engineering, University of Western Macedonia, Kozani 50100, Greece.\\
\IEEEauthorrefmark{2} Research \& Development Department, InnoCube P.C., Thessaloniki 55535, Greece.\\
\IEEEauthorrefmark{3} Department of Electronics and Communication Engineering,  Manipal Institute of Technology,\\ Manipal Academy of Higher Education, Manipal 576104, India. \\
\IEEEauthorrefmark{4} Department of Informatics \& Telecommunications, University of Thessaly, Lamia 35100, Greece.\\
\IEEEauthorrefmark{5} Department of Electrical and Electronic Engineering, University of Nottingham Ningbo China, Ningbo 315100, \\ China.
\IEEEauthorblockA{Emails: 
\{dece00106, aboulogeorgos\}@uowm.gr, 
trevlakis@innocube.org, 
tanweer.ali@manipal.edu, 
tsiftsis@uth.gr}
}   


}
\maketitle	
\begin{abstract}
While the wireless word moves towards higher frequency bands, new challenges arises, due to the inherent characteristics of the transmission links, such as high path and penetration losses. Penetration losses causes blockages that in turn can significantly reduce the signal strength at the receiver. Most published contributions consider a binary blockage stage, i.e. either fully blocked or blockage-free links. However, in realistic scenarios, a link can be partially blocked. Motivated by this, in this paper, we present two low-complexity models that are based on tight approximations and accommodates the impact of partial blockage in high-frequency links. To demonstrate the applicability of the derived framework, we present closed-form expressions for the outage probability for the case in which the distance between the center of the receiver plane and the blocker's shadow center follow uniform distribution. Numerical results verify the derived framework and reveal how the transmission parameters affect blockage.           
\end{abstract}

\section{Introduction}\label{S:Intro}

As we move towards the next generation wireless systems and networks, the telecommunication traffic in the network is expected to exponentially increase, due to the development of killer-applications, such as extended reality, three-dimensional printing, digital twins, etc., with significant demands on data-rate and inherent security~\cite{10302317}. This creates the need for employing underutilized and non-standardized frequency bands to deal with spectrum scarcity becomes more and more prominent~\cite{9583918,9615497,9356523}. As a consequence, the wireless world turned its attention to millimeter wave (mmW), sub-terahetz (THz), THz, and optical bands~\cite{Boulogeorgos2018,10528305}. However, links that are established in the aforementioned bands suffer from high penetration losses that result in blockages.   

Scanning the technical literature, there are several contributions that model and quantify the impact of blockage~\cite{Kizhakkekundil2021, Alsaleem2020, Tang2020, Gapeyenko2020,Wu2021,Wang2023}. The authors of~\cite{Kizhakkekundil2021} assumed the obstacles as knife-edge  rectangular shapes and categorize them at three categories depending on their dimensions. In the same direction, the authors of~\cite{Alsaleem2020} categorized obstacles into four- and two-edge obstacles.  In ~\cite{Tang2020}, the authors examined the impact of blockage in an indoor wireless network assisted by unmanned aerial vehicles. The authors of ~\cite{Gapeyenko2020} studied the effect of human body blockage and introduced a mathematical framework to quantify its impact. The aforementioned contributions focus on obstacle dimensions and their proposed models based on the third-generation partnership project (3GPP) blockage model ~\cite{3GPP2024}. All of the presented techniques focus on overcoming the blockage impact, and none of them on understanding its impact on the received signal. 

Inspired by the above fact, in~\cite{Wu2021}, the authors employed stochastic geometry to analyze the effect of blockage in indoor THz wireless systems. The analysis of~\cite{Wu2021} was based on modeling the blockers as cylinders of a randomly chosen position and height, and assuming that the link is either free of blockage or fully blocked. A similar approach was followed in~\cite{Wang2023}, where the authors assumed an exponential random variable to model the probability of establishing a blockage-free link. Again, no partially blocked links were considered.   

 To cover this research gap, in this paper, we introduce a novel blockage model that accounts for partial blockage. In more detail, we present two low-complexity approximated expressions for the impact of blockage coefficient and we verify their accuracy after analytically comparing them with the exact expressions in terms of mean square error (MSE) and normalized MSE (NMSE). In order to show the applicability of the approximations in complex environments, we present the outage probability of a wireless link that suffer from partial blockage, for the case in which the distance between the center of the receiver plane and the blocker's shadow center follow uniform distribution. The methodology that is adopted for this contribution can be generalized for any distribution of the corresponding distance. 


\subsubsection*{Notations} 
The absolute value, exponential and natural logarithm functions are respectively denoted by $|\cdot|$,  $\exp\left(\cdots\right)$, and $\ln\left(\cdot\right)$.
  $\sqrt{x}$ and $\prod_{l=1}^{L}x_l$ respectively return the square root of $x$,  and the product of $x_1\,x_2\,\cdots\, x_L$.  
 $\Pr\left(\mathcal{A}\right)$ denotes the probability for the event $\mathcal{A}$ to be valid. 
The modified Bessel function of the second kind of order $n$ is denoted as~$\mathrm{K}_n(\cdot)$~\cite[eq. (8.407/1)]{B:Gra_Ryz_Book}. 
The  Gamma~\cite[eq. (8.310)]{B:Gra_Ryz_Book} function is  denoted by  $\Gamma\left(\cdot\right)$, and the error-function is represented by $\erf\left(\cdot\right)$~\cite[eq. (8.250/1)]{B:Gra_Ryz_Book}. Finally,  $G_{p, q}^{m, n}\left(x\left| \begin{array}{c} a_1, a_2, \cdots, a_{p} \\ b_{1}, b_2, \cdots, b_q\end{array}\right.\right)$  stands for the Meijer G-function~\cite[eq. (9.301)]{B:Gra_Ryz_Book}.

\section{Blockage Characterization}\label{S:BC}

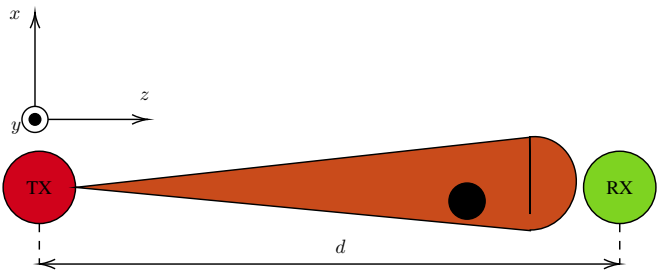
\begin{figure}
	\centering
	\scalebox{.73}{
		
		\tikzset{every picture/.style={line width=0.75pt}} 
		
		\begin{tikzpicture}[x=0.75pt,y=0.75pt,yscale=-1,xscale=1]
			
			\draw  [fill={rgb, 255:red, 208; green, 2; blue, 27 }  ,fill opacity=0.5 ] (23,130) .. controls (23,116.19) and (34.19,105) .. (48,105) .. controls (61.81,105) and (73,116.19) .. (73,130) .. controls (73,143.81) and (61.81,155) .. (48,155) .. controls (34.19,155) and (23,143.81) .. (23,130) -- cycle ;
			\draw  [fill={rgb, 255:red, 126; green, 211; blue, 33 }  ,fill opacity=0.5 ] (424,130) .. controls (424,116.19) and (435.19,105) .. (449,105) .. controls (462.81,105) and (474,116.19) .. (474,130) .. controls (474,143.81) and (462.81,155) .. (449,155) .. controls (435.19,155) and (424,143.81) .. (424,130) -- cycle ;
			\draw  [draw opacity=0][fill={rgb, 255:red, 201; green, 75; blue, 27 }  ,fill opacity=0.5 ][line width=0.75]  (386.74,95.36) -- (387.14,148.43) -- (386.79,95.11) .. controls (390.47,94.66) and (394.25,94.93) .. (397.98,96.02) .. controls (413.83,100.65) and (422.61,118.35) .. (417.58,135.56) .. controls (413.34,150.08) and (400.62,159.61) .. (387.22,159.44) -- (387.23,159.92) -- (73,130) -- (386.74,95.36) -- cycle ;
			\draw  [fill={rgb, 255:red, 0; green, 0; blue, 0 }  ,fill opacity=1 ] (331,139.5) .. controls (331,132.6) and (336.6,127) .. (343.5,127) .. controls (350.4,127) and (356,132.6) .. (356,139.5) .. controls (356,146.4) and (350.4,152) .. (343.5,152) .. controls (336.6,152) and (331,146.4) .. (331,139.5) -- cycle ;
			\draw    (54,83) -- (121,83) ;
			\draw [shift={(123,83)}, rotate = 180] [color={rgb, 255:red, 0; green, 0; blue, 0 }  ][line width=0.75]    (10.93,-3.29) .. controls (6.95,-1.4) and (3.31,-0.3) .. (0,0) .. controls (3.31,0.3) and (6.95,1.4) .. (10.93,3.29)   ;
			\draw   (36,83) .. controls (36,78.03) and (40.03,74) .. (45,74) .. controls (49.97,74) and (54,78.03) .. (54,83) .. controls (54,87.97) and (49.97,92) .. (45,92) .. controls (40.03,92) and (36,87.97) .. (36,83) -- cycle ;
			\draw    (45,74) -- (45,11) ;
			\draw [shift={(45,9)}, rotate = 450] [color={rgb, 255:red, 0; green, 0; blue, 0 }  ][line width=0.75]    (10.93,-3.29) .. controls (6.95,-1.4) and (3.31,-0.3) .. (0,0) .. controls (3.31,0.3) and (6.95,1.4) .. (10.93,3.29)   ;
			\draw  [fill={rgb, 255:red, 0; green, 0; blue, 0 }  ,fill opacity=1 ] (41,83) .. controls (41,80.79) and (42.79,79) .. (45,79) .. controls (47.21,79) and (49,80.79) .. (49,83) .. controls (49,85.21) and (47.21,87) .. (45,87) .. controls (42.79,87) and (41,85.21) .. (41,83) -- cycle ;
			\draw  [dash pattern={on 4.5pt off 4.5pt}]  (48,155) -- (48,183) ;
			\draw  [dash pattern={on 4.5pt off 4.5pt}]  (449,155) -- (449,183) ;
			\draw    (50,183) -- (447,183) ;
			\draw [shift={(449,183)}, rotate = 180] [color={rgb, 255:red, 0; green, 0; blue, 0 }  ][line width=0.75]    (10.93,-3.29) .. controls (6.95,-1.4) and (3.31,-0.3) .. (0,0) .. controls (3.31,0.3) and (6.95,1.4) .. (10.93,3.29)   ;
			\draw [shift={(48,183)}, rotate = 0] [color={rgb, 255:red, 0; green, 0; blue, 0 }  ][line width=0.75]    (10.93,-3.29) .. controls (6.95,-1.4) and (3.31,-0.3) .. (0,0) .. controls (3.31,0.3) and (6.95,1.4) .. (10.93,3.29)   ;
			
			\draw (48,130) node   [align=left] {TX};
			\draw (449,130) node   [align=left] {RX};
			\draw (116,62.4) node [anchor=north west][inner sep=0.75pt]    {$z$};
			\draw (26,6.4) node [anchor=north west][inner sep=0.75pt]    {$x$};
			\draw (27,83.4) node [anchor=north west][inner sep=0.75pt]    {$y$};
			\draw (251,164.4) node [anchor=north west][inner sep=0.75pt]    {$d$};

		\end{tikzpicture}
		
		}
	\caption{System model.}
	\label{Fig:SM}
\end{figure}

As demonstrated in Fig.~\ref{Fig:SM}, we consider a wireless setup that consists of one transmitter (TX), one receiver (RX) and a single obstacle, which is located between the TX and RX. Moreover, we assume that the TX-RX link is highly directional. Note that this is the usual case in high-frequency communications such as millimeter waves (mmW) and terahertz (THz). In this band, the Gaussian TX beam approximation is considered very accurately~\cite{A:Analytical_Performance_Assessment_of_THz_Wireless_Systems}. Thus, the normalized spatial distribution of the transmitted intensity at distance $d$ from the TX can be obtained as~\cite{Farid2007}
\begin{align}
	U_b\left(\mathbold{\rho}, d\right) = \frac{2}{\pi w_{d}^2} \exp\left(-\frac{2\left|\left|\mathbold{\rho} \right|\right|^{2}}{w_d^2}\right),
\end{align}
where $\mathbold{\rho}$ stands for the radial vector from the beam center at distance $d$. Moreover, $w_d$ represents the beam waste at distance~$d$. 

By considering a circular detection aperture of radius $\alpha$ at the RX, we can evaluate the geometric spread~as
\begin{align}
	h_b\left(\mathbold{r}; d\right) = \int_{\mathcal{A}} U_b\left(\mathbold{\rho}, d\right)\,\mathrm{d}\mathbold{\rho} - \int_{\mathcal{A}_b} U_b\left(\mathbold{\rho}-\mathbold{r}, d\right)\,\mathrm{d}\mathbold{\rho},
	\label{Eq:h_p_1}
\end{align} 
where $h_b\left(\cdot; \cdot\right)$ represents the fraction of power collected by the RX. Furthermore, $\mathcal{A}$ and $\mathcal{A}_b$, respectively, stand for the effective areas of RX and the areas shadowed by the blocker areas at the RX~plane. 

\begin{figure}
	\centering
	\scalebox{.73}{

	\tikzset{every picture/.style={line width=0.75pt}} 
	
	\begin{tikzpicture}[x=0.75pt,y=0.75pt,yscale=-1,xscale=1]
		
		\draw    (201,169) -- (201,28) ;
		\draw [shift={(201,26)}, rotate = 450] [color={rgb, 255:red, 0; green, 0; blue, 0 }  ][line width=0.75]    (10.93,-3.29) .. controls (6.95,-1.4) and (3.31,-0.3) .. (0,0) .. controls (3.31,0.3) and (6.95,1.4) .. (10.93,3.29)   ;
		\draw    (201,169) -- (366,169) ;
		\draw [shift={(368,169)}, rotate = 180] [color={rgb, 255:red, 0; green, 0; blue, 0 }  ][line width=0.75]    (10.93,-3.29) .. controls (6.95,-1.4) and (3.31,-0.3) .. (0,0) .. controls (3.31,0.3) and (6.95,1.4) .. (10.93,3.29)   ;
		\draw  [fill={rgb, 255:red, 201; green, 75; blue, 27 }  ,fill opacity=0.5 ] (103.5,169) .. controls (103.5,115.15) and (147.15,71.5) .. (201,71.5) .. controls (254.85,71.5) and (298.5,115.15) .. (298.5,169) .. controls (298.5,222.85) and (254.85,266.5) .. (201,266.5) .. controls (147.15,266.5) and (103.5,222.85) .. (103.5,169) -- cycle ;
		\draw  [fill={rgb, 255:red, 126; green, 211; blue, 33 }  ,fill opacity=0.5 ] (143.25,169) .. controls (143.25,137.11) and (169.11,111.25) .. (201,111.25) .. controls (232.89,111.25) and (258.75,137.11) .. (258.75,169) .. controls (258.75,200.89) and (232.89,226.75) .. (201,226.75) .. controls (169.11,226.75) and (143.25,200.89) .. (143.25,169) -- cycle ;
		\draw  [fill={rgb, 255:red, 0; green, 0; blue, 0 }  ,fill opacity=0.5 ] (222.88,169) .. controls (222.88,149.19) and (238.94,133.13) .. (258.75,133.13) .. controls (278.56,133.13) and (294.63,149.19) .. (294.63,169) .. controls (294.63,188.81) and (278.56,204.88) .. (258.75,204.88) .. controls (238.94,204.88) and (222.88,188.81) .. (222.88,169) -- cycle ;
		\draw [color={rgb, 255:red, 255; green, 255; blue, 255 }  ,draw opacity=1 ]   (251,144) -- (224,159) ;
		\draw [color={rgb, 255:red, 255; green, 255; blue, 255 }  ,draw opacity=1 ]   (257,151) -- (222.88,169) ;
		\draw [color={rgb, 255:red, 255; green, 255; blue, 255 }  ,draw opacity=1 ]   (260,162) -- (225.88,180) ;
		\draw [color={rgb, 255:red, 255; green, 255; blue, 255 }  ,draw opacity=1 ]   (259,174) -- (232,189) ;
		\draw [color={rgb, 255:red, 255; green, 255; blue, 255 }  ,draw opacity=1 ]   (258,185) -- (238,197) ;
		\draw [color={rgb, 255:red, 255; green, 255; blue, 255 }  ,draw opacity=1 ]   (248,138) -- (229,148) ;
		\draw [color={rgb, 255:red, 255; green, 255; blue, 255 }  ,draw opacity=1 ][line width=1.5]    (258.75,169) -- (289,151) ;
		\draw    (312,236) .. controls (262.75,228.12) and (266.86,236.73) .. (250.75,204.5) ;
		\draw [shift={(250,203)}, rotate = 423.43] [color={rgb, 255:red, 0; green, 0; blue, 0 }  ][line width=0.75]    (10.93,-3.29) .. controls (6.95,-1.4) and (3.31,-0.3) .. (0,0) .. controls (3.31,0.3) and (6.95,1.4) .. (10.93,3.29)   ;
		\draw    (83,243) .. controls (125.36,243) and (134.72,240.09) .. (156.02,212.29) ;
		\draw [shift={(157,211)}, rotate = 487.18] [color={rgb, 255:red, 0; green, 0; blue, 0 }  ][line width=0.75]    (10.93,-3.29) .. controls (6.95,-1.4) and (3.31,-0.3) .. (0,0) .. controls (3.31,0.3) and (6.95,1.4) .. (10.93,3.29)   ;
		\draw [color={rgb, 255:red, 0; green, 0; blue, 0 }  ,draw opacity=1 ][line width=1.5]    (157,137) -- (201,169) ;
		\draw  [dash pattern={on 0.84pt off 2.51pt}]  (320,73) .. controls (278.72,70.12) and (276.15,79.22) .. (264.51,89.69) ;
		\draw [shift={(263,91)}, rotate = 319.76] [color={rgb, 255:red, 0; green, 0; blue, 0 }  ][line width=0.75]    (10.93,-3.29) .. controls (6.95,-1.4) and (3.31,-0.3) .. (0,0) .. controls (3.31,0.3) and (6.95,1.4) .. (10.93,3.29)   ;
		\draw  [dash pattern={on 0.84pt off 2.51pt}]  (49,148) .. controls (50.95,171.4) and (114.69,171.03) .. (141.28,169.15) ;
		\draw [shift={(143.25,169)}, rotate = 535.47] [color={rgb, 255:red, 0; green, 0; blue, 0 }  ][line width=0.75]    (10.93,-3.29) .. controls (6.95,-1.4) and (3.31,-0.3) .. (0,0) .. controls (3.31,0.3) and (6.95,1.4) .. (10.93,3.29)   ;
		\draw  [dash pattern={on 0.84pt off 2.51pt}]  (357,207) .. controls (315.72,207) and (305.77,198.71) .. (293.54,189.19) ;
		\draw [shift={(292,188)}, rotate = 397.57] [color={rgb, 255:red, 0; green, 0; blue, 0 }  ][line width=0.75]    (10.93,-3.29) .. controls (6.95,-1.4) and (3.31,-0.3) .. (0,0) .. controls (3.31,0.3) and (6.95,1.4) .. (10.93,3.29)   ;
		\draw  [dash pattern={on 4.5pt off 4.5pt}]  (258.75,41) -- (258.75,169) ;
		\draw    (256.75,41) -- (203,41) ;
		\draw [shift={(201,41)}, rotate = 360] [color={rgb, 255:red, 0; green, 0; blue, 0 }  ][line width=0.75]    (10.93,-3.29) .. controls (6.95,-1.4) and (3.31,-0.3) .. (0,0) .. controls (3.31,0.3) and (6.95,1.4) .. (10.93,3.29)   ;
		\draw [shift={(258.75,41)}, rotate = 180] [color={rgb, 255:red, 0; green, 0; blue, 0 }  ][line width=0.75]    (10.93,-3.29) .. controls (6.95,-1.4) and (3.31,-0.3) .. (0,0) .. controls (3.31,0.3) and (6.95,1.4) .. (10.93,3.29)   ;
		
		\draw (270.27,148.05) node  [color={rgb, 255:red, 255; green, 255; blue, 255 }  ,opacity=1 ,rotate=-330.67]  {$\alpha _{b}$};
		\draw (324.41,239.05) node    {$\mathcal{A}_{b}$};
		\draw (72.41,246.05) node    {$\mathcal{A}$};
		\draw (183.35,132.32) node [anchor=north west][inner sep=0.75pt]  [rotate=-40.35]  {$\alpha $};
		\draw (355.43,58) node [anchor=north] [inner sep=0.75pt]   [align=left] {\begin{minipage}[lt]{44.1pt}\setlength\topsep{0pt}
				\begin{center}
					TX beam\\footprint
				\end{center}
				
		\end{minipage}};
		\draw (48.43,108) node [anchor=north] [inner sep=0.75pt]   [align=left] {\begin{minipage}[lt]{57.53pt}\setlength\topsep{0pt}
				\begin{center}
					RX effective\\area
				\end{center}
				
		\end{minipage}};
		\draw (386.43,193) node [anchor=north] [inner sep=0.75pt]   [align=left] {\begin{minipage}[lt]{37.88pt}\setlength\topsep{0pt}
				\begin{center}
					Blocker\\shadow
				\end{center}
				
		\end{minipage}};
		\draw (378.22,167) node    {$x$};
		\draw (201.22,12) node    {$y$};
		\draw (231.22,31) node    {$r$};

	\end{tikzpicture}
	}
	\caption{RX plane.}
	\label{Fig:RX_plane}
\end{figure}
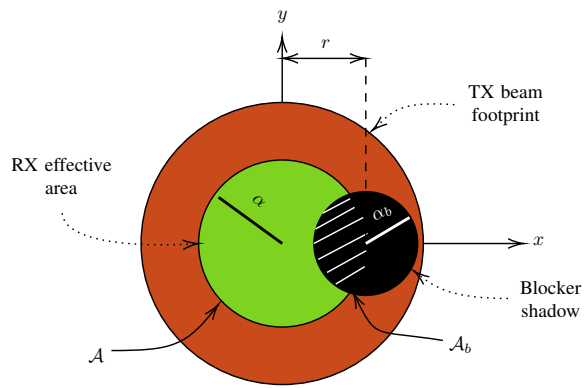

For tractability, we employ the well-used round-ball approximation for the blocker. In general, when the center of the blocker's shadow at the RX plane is at $\mathbold{r}$, $h_b$ is a function of $\left|\left|\mathbold{r}\right|\right|$ as well as the corresponding angle. However, due to the symmetry of the beam and blocker shapes as well as the effective area, $h_b\left(\mathbold{r}; d\right)$ depends only on $r=\left|\left|\mathbold{r}\right|\right|$. Hence, as depicted in Fig.~\ref{Fig:RX_plane}, we can assume that the blocker shadow is located along $x-$axis. Thus,~\eqref{Eq:h_p_1} can be rewritten~as 
\begin{align}
	h_b\left(\mathbold{r}; d\right) = \mathcal{I} - \mathcal{I}_b,
\end{align} 
where
\begin{align}
	\mathcal{I} = \int_{-\alpha}^{\alpha} \int_{-\alpha}^{\alpha} \frac{2}{\pi w_{d}^2} \exp\left(-2 \frac{x^2+y^2}{w_d^2} \right)\,\mathrm{d}x \,\mathrm{d}y 
	\label{Eq:I}
\end{align} 
and 
\begin{align}
	\mathcal{I}_b = \int_{-\alpha_b}^{\alpha_b} \int_{-\zeta}^{\zeta} \frac{2}{\pi w_{d}^2} \exp\left(-2 \frac{(x-r)^2+y^2}{w_d^2} \right)\,\mathrm{d}y \,\mathrm{d}x, 
	\label{Eq:I_b}
\end{align}  
with 
\begin{align}
	\zeta = \sqrt{a_b^2 - x^2}.
	\label{Eq:zeta} 
\end{align}
Moreover, note that in order for the obstacle to influence the portion of reception power, the following inequality should~hold: 
\begin{align}
	r-a_b<a.
\end{align}

After some algebraic manipulations,~\eqref{Eq:I} can be rewritten~as
\begin{align}
	\mathcal{I} = \frac{2}{\pi w_{d}^2} \int_{-\alpha}^{\alpha} \exp\left(-2 \frac{y^2}{w_d^2} \right) \mathrm{d}y \, \int_{-\alpha}^{\alpha}  \exp\left(-2 \frac{x^2}{w_d^2} \right)\,\mathrm{d}x,
	\label{Eq:I_s2}
\end{align} 
or equivalently 
\begin{align}
	\mathcal{I} = \left(\erf\left(\frac{\sqrt{2} \alpha}{w_d}\right)\right)^2.
	\label{Eq:I_s3}
\end{align}

Similarly,~\eqref{Eq:I_b} can be rewritten~as
\begin{align}
		\mathcal{I}_b = \frac{2}{\pi w_d^2} \int_{-\alpha_b}^{\alpha_b} \exp\left(-2\frac{(x-r)^2}{w_d^2}\right) \mathcal{K}(\zeta)\, \mathrm{d}x,
		\label{Eq:Ib_s2}
\end{align}
where
\begin{align}
	\mathcal{K}(\zeta) = \int_{-\zeta}^{\zeta}\exp\left(-2\frac{y^2}{w_d^2}\right)\,\mathrm{d}y,
\end{align}
which can be rewritten~as
\begin{align}
	\mathcal{K}(\zeta) = \sqrt{\frac{\pi}{2}} w_d \erf\left(\frac{\sqrt{2} \zeta}{w_d} \right).
	\label{Eq:K}
\end{align}
By applying~\eqref{Eq:zeta} and~\eqref{Eq:K} into~\eqref{Eq:Ib_s2}, we get
\begin{align}
	\mathcal{I}_b = \sqrt{\frac{2}{\pi}} \frac{1}{w_d} &\int_{-\alpha_b}^{\alpha_b}  \exp\left(-2\frac{(x-r)^2}{w_d^2}\right) 
	\nonumber \\ & \times 	
	\erf\left(\frac{\sqrt{2} \sqrt{a_b^2 - x^2}}{w_d} \right)\, \mathrm{d}x.
	\label{Eq:I_b_exact}
\end{align}

Unfortunately, it is very difficult or even impossible to write~\eqref{Eq:I_b_exact} in a closed form. The following theorems return closed-form tight approximations for~$I_b$.

\begin{thm}
	A tight closed-form approximation for~\eqref{Eq:Ib_s2} can be obtained~as
	\begin{align}
		\mathcal{I}_b &\approx \frac{1}{2} \erf\left(\frac{a_b\sqrt{\pi}}{\sqrt{2} w_d}\right)
		\nonumber \\ & \times
		\left( \erf\left( \frac{a_b \sqrt{\pi} - 2 r}{\sqrt{2} w_d} \right) + \erf\left( \frac{a_b \sqrt{\pi} + 2 r}{\sqrt{2} w_d} \right) \right).
		\label{Eq:I_b_approx_s2} 
	\end{align}
\end{thm}
\begin{IEEEproof}
 By considering approximation the integration in~\eqref{Eq:I_b} by an integration over a square of equal area to the obstacle, i.e., with side length $\sqrt{\pi}\alpha_b$,~\eqref{Eq:I_b} can be approximated~as    
\begin{align}
	\mathcal{I}_b \approx \frac{2}{\pi w_d^2} \int_{-\frac{\sqrt{\pi}a_b}{2}}^{\frac{\sqrt{\pi}a_b}{2}} &\int_{-\frac{\sqrt{\pi}a_b}{2}}^{\frac{\sqrt{\pi}a_b}{2}} \exp\left(-2\frac{\left(x-r\right)^2}{w_d^2}\right) 
	\nonumber \\ & \times
	\exp\left(-2\frac{y^2}{w_d^2}\right) \mathrm{d}x \, \mathrm{d}y, 
	\label{Eq:I_b_approx}
\end{align}
which can be written in a closed-form~as in~\eqref{Eq:I_b_approx_s2}. This concludes the proof.
\end{IEEEproof}

\begin{figure}
	\centering\includegraphics[width=1\linewidth,trim=0 0 0 0,clip=false]{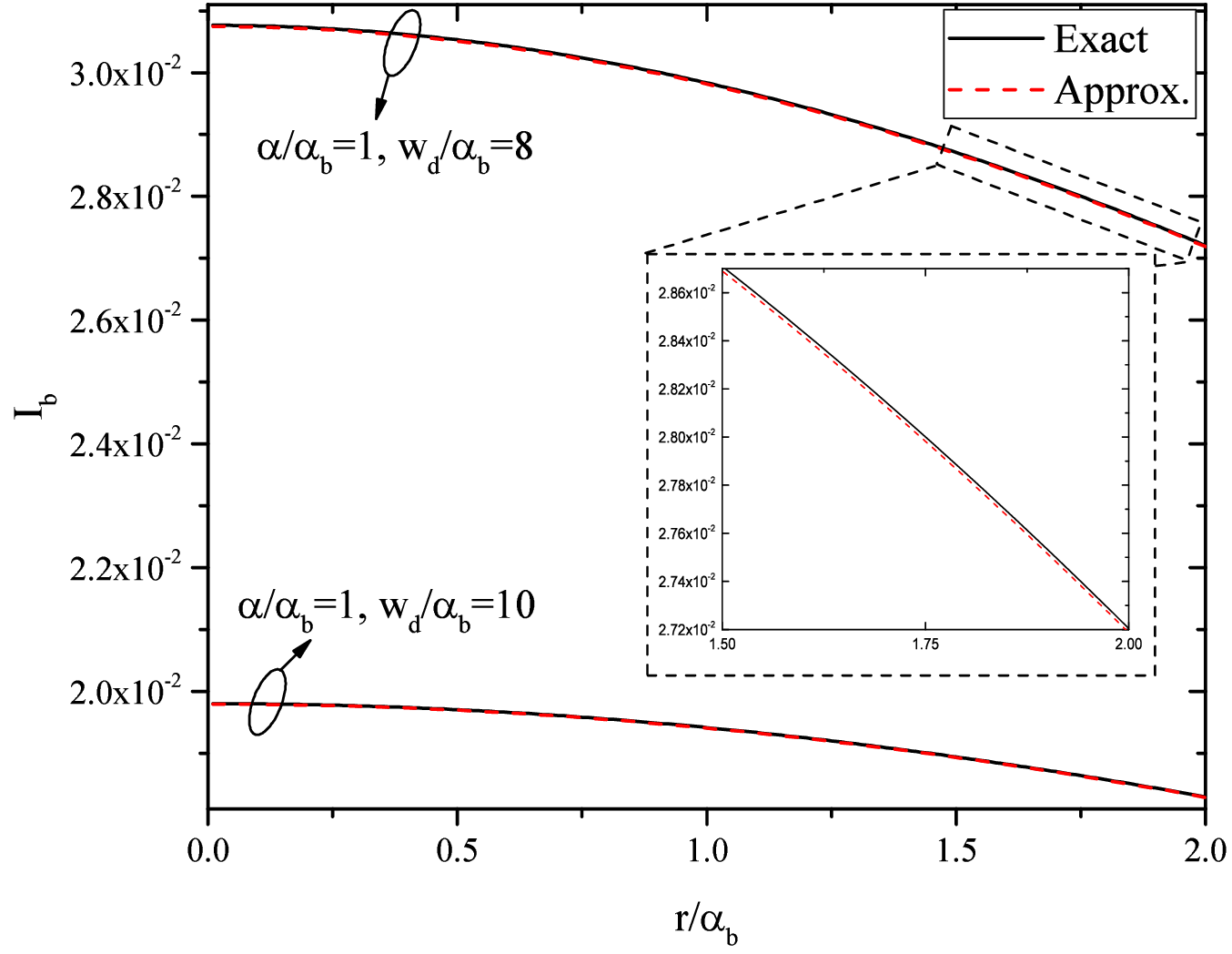}
	\caption{Exact and approximate values of $I_b$ as a function of $r/\alpha_b$, for different values of $w_d/\alpha_b$.}\label{fig:SM}
	\label{Fig:Approx}
\end{figure}

The approximation presented in Theorem 1, although very tight, does not provide a tractable solution for performance analysis. Motivated by this, the following theorem returns a simplified approximation of $I_b$. 
\begin{thm}
	A simple closed-form approximation for~\eqref{Eq:I_b_exact} can be obtained~as
	\begin{align}
		\mathcal{I}_b \approx C_0 \exp\left(-\frac{2}{w_d^2} \frac{\left|C_2\right|}{C_0} r^2 \right),
		\label{Eq:I_b_approx_2}
	\end{align}
	where 
	\begin{align}
		C_0 = \left( \erf\left(\sqrt{\frac{\pi}{2}} \frac{\alpha_b}{w_d} \right) \right)^2
	\end{align}
	and
	\begin{align}
		C_2 = - \sqrt{2} \frac{\alpha_b}{w_d}
		\erf{\left(\sqrt{\frac{\pi}{2}} \frac{\alpha_b}{w_d} \right)} 
		\exp\left(-\left(\sqrt{\frac{\pi}{2}} \frac{\alpha_b}{w_d} \right)^2\right).
	\end{align}
\end{thm}

\begin{IEEEproof}
First, we rewrite~\eqref{Eq:I_b_approx}~as
\begin{align}
		\mathcal{I}_b \approx  \frac{2}{\pi w_d^2} \mathcal{J} \int_{-\frac{\sqrt{\pi}a_b}{2}}^{\frac{\sqrt{\pi}a_b}{2}}  \exp\left(-2\frac{\left(x-r\right)^2}{w_d^2}\right) 
	 \mathrm{d}x, 
	\label{Eq:I_b_approx_al_s1}
\end{align}
where 
\begin{align}
	\mathcal{J} = \int_{-\frac{\sqrt{\pi}a_b}{2}}^{\frac{\sqrt{\pi}a_b}{2}}  \exp\left(-2\frac{y^2}{w_d^2}\right) 
	\mathrm{d}y,
\end{align}
which can be expressed in closed-form~as
\begin{align}
	\mathcal{J} = \sqrt{\frac{\pi}{2}} w_d \erf\left( \sqrt{\frac{\pi}{2}} \frac{a_b}{w_d}\right).
\end{align}
By using the Taylor series to extend the exponential term in~\eqref{Eq:I_b_approx_al_s1}, we get
 \begin{align}
 	\mathcal{I}_b \approx  \frac{2\alpha_b}{ w_d} \mathcal{J} &+ \frac{2}{\sqrt{\pi}}  \mathcal{J} \sum_{\begin{array}{c} k= 3 \\ \text{odd} \end{array}}^{\infty} \sum_{\begin{array}{c} l= 0 \\ \text{even} \end{array}}^{k}\frac{\left(-1\right)^{(k-1)/2}}{k \left(\frac{k-1}{2}\right)!}
 	\nonumber \\ & \times
 	\left(\begin{array}{c} m \\ l \end{array} \right) \left(\sqrt{2} \frac{r}{w_d}\right)^{l} \left(\sqrt{\frac{\pi}{2}} \frac{\alpha_b}{w_d}\right)^{k-l},
 \end{align}
which can be rewritten~as
\begin{align}
	\mathcal{I}_b \approx  \sum_{\begin{array}{c} l= 0 \\ \text{even} \end{array}}^{\infty} C_l \left(\frac{\sqrt{2}}{w_d} r \right)^l,
	\label{Eq:I_b_approx_al_s2}
\end{align}
where 
\begin{align}
	C_l = \frac{2}{\sqrt{\pi}} \mathcal{J} \sum_{\begin{array}{c} k=l+1\\ \text{odd}\end{array}}^{\infty} &
	 \frac{\left(-1\right)^{(k-1)/2}}{k \left(\frac{k-1}{2}\right)!} \left(\begin{array}{c} m \\ l \end{array} \right) \left(\sqrt{2} \frac{r}{w_d}\right)^{l} 
	 \nonumber \\ & \times
	 \left(\sqrt{\frac{\pi}{2}} \frac{\alpha_b}{w_d}\right)^{k-l}.
\end{align}
By equating the first two terms of the Taylor expansion of the Gaussian pulse to the same terms of~\eqref{Eq:I_b_approx_al_s2}, we get~\eqref{Eq:I_b_approx_2}. This concludes the proof.
\end{IEEEproof}

Figure~\ref{Fig:Approx} demonstrates the precision of the approximation presented in~\eqref{Eq:I_b_approx_s2}. In more detail, the exact and approximate values of $I_b$ are plotted against of $r/\alpha_b$, for different values of $w_d/\alpha_b$. From this figure, it becomes evident that the maximum approximation error is in the order of $10^{-5}$. Similarly, Table~\ref{T:Approx_error} presents the mean square error (MSE) and normalized MSE (NMSE) between the exact and simplified approximations of $I_b$ expressions for different values of $w_d/\alpha_b$. From this table, we observe that both MSE and NMSE are relatively small; thus, the approximation can be considered to be tight.

	\begin{table}
		\caption{MSE and NMSE between exact and simplified approximate $I_b$ expressions.}
		\label{T:Approx_error}
		\centering 
	\begin{tabular}{ |c|c|c| } 
		\hline
		$w_d/\alpha_{b}$ & MSE & NMSE \\ 
		\hline \hline
		$2$ & $5.81\times 10^{-6}$ & $5.99\times 10^{-5}$ \\ 
		\hline
		$3$ & $3.29\times 10^{-7}$ & $1.52\times 10^{-5}$ \\
		\hline
		$4$ & $3.73\times 10^{-8}$ & $5.25\times 10^{-6}$ \\
		\hline
		$5$ & $6.59\times 10^{-9}$ & $2.25\times 10^{-6}$ \\
		\hline
		$6$ & $1.59\times 10^{-9}$ & $1.11\times 10^{-6}$ \\
		\hline
		$7$ & $4.65\times 10^{-10}$ & $6.07\times 10^{-7}$ \\
		\hline
		$8$ & $1.59\times 10^{-10}$ & $3.59\times 10^{-7}$ \\
		\hline
		$9$ & $6.3\times 10^{-11}$ & $2.25\times 10^{-7}$ \\
		\hline
		$10$ & $2.69\times 10^{-11}$ & $1.49\times 10^{-7}$ \\
		\hline
	\end{tabular}
	\end{table}

Next, applying~\eqref{Eq:I_s3} and~\eqref{Eq:I_b_approx_s2} to~\eqref{Eq:h_p_1}, we obtain~\eqref{Eq:h_b_s2}, given at the top of the next page. 
\begin{figure*}
	\begin{align}
		h_b^{(1)}\left(\mathbold{r}; d\right) \approx  \left(\erf\left(\frac{\sqrt{2} \alpha}{w_d}\right)\right)^2 - \frac{1}{2} \erf\left(\frac{a_b\sqrt{\pi}}{\sqrt{2} w_d}\right)
		\left( \erf\left( \frac{a_b \sqrt{\pi} - 2 r}{\sqrt{2} w_d} \right) + \erf\left( \frac{a_b \sqrt{\pi} + 2 r}{\sqrt{2} w_d} \right) \right)
		\label{Eq:h_b_s2}
	\end{align}
	\hrulefill
\end{figure*}
Moreover, by applying~\eqref{Eq:I_s3} and~\eqref{Eq:I_b_approx_2} into~\eqref{Eq:h_p_1}, we get~
	\begin{align}
		h_b^{(2)}\left(\mathbold{r}; d\right) \approx  \left(\erf\left(\frac{\sqrt{2} \alpha}{w_d}\right)\right)^2 -  C_0 \exp\left(-\frac{2}{w_d^2} \frac{\left|C_2\right|}{C_0} r^2 \right).
		\label{Eq:h_b_2}
	\end{align}

 From~\eqref{Eq:h_b_2}, it becomes evident that
 \begin{align}
     0 \leq h_b^{(2)} \leq 1,
 \end{align}
 or equivalently
 \begin{align}
     0 \leq \left(\erf\left(\frac{\sqrt{2} \alpha}{w_d}\right)\right)^2 -  C_0 \exp\left(-\frac{2}{w_d^2} \frac{\left|C_2\right|}{C_0} r^2 \right) \leq 1,
 \end{align}
 which leads to the following inequality
 \begin{align}
     A_1
      \leq r \leq 
      A_2
     \label{Eq:ineq}
 \end{align}
where
\begin{align}
    A_1 = \frac{w_d}{\sqrt{2}} \, \sqrt{\frac{C_0}{\left|C_2\right|}} \, \sqrt{\ln\left(\frac{C_0}{\left(\erf\left(\frac{\sqrt{2} \alpha}{w_d}\right)\right)^2}\right)} 
\end{align}
and
\begin{align}
    A_2 = \frac{w_d}{\sqrt{2}} \, \sqrt{\frac{C_0}{\left|C_2\right|}} \, \sqrt{\ln\left(\frac{C_0}{\left(\erf\left(\frac{\sqrt{2} \alpha}{w_d}\right)\right)^2 - 1}\right)}.
\end{align}

\section{Applications}

Let us assume that $r$ follows a uniform distribution with probability distribution function (PDF) and cumulative density function (CDF) that can be respectively obtained~as in~\cite{papoulis}
\begin{align}
    f_{r}(x) = \left\{\begin{array}{l l} \frac{1}{A_2-A_1}, & \text{for } A_1 \leq x \leq A_2 \\ 0, & \text{otherwise} \end{array}\right.
\end{align}
and 
\begin{align}
    F_{r}(x) = \left\{
                        \begin{array}{l l} 
                            0, & \text{for } x < A_1 \\ 
                            \frac{1}{A_2-A_1}\left(x-A_1\right) & \text{for }  A_1 \leq x \leq A_2 \\
                            1, & \text{for } x > A_2
                        \end{array}
                \right.
                \label{Eq:Fr}
\end{align}

The following proposition returns a closed-form expression for the outage probability of the link, in the case $r$ follows a uniform distribution. 
\begin{prop}
    If $r$ follows a uniform distribution, the outage probability of the link can be obtained~as in~\eqref{Eq:P_o_final}, given at the top of the next page. 
    \begin{figure*}
    \begin{align}
        P_o = \left\{
                        \begin{array}{l l} 
                            0, &  \text{for } \mathcal{C}_1  \\ 
                            \frac{1}{A_2-A_1}\left(\frac{w_d}{\sqrt{2}} \sqrt{\frac{C_0}{\left|C_2\right|}}\sqrt{\ln\left(\frac{1}{C_0}\left(\erf\left(\frac{\sqrt{2} \alpha}{w_d}\right)\right)^2 + \frac{1}{C_0} \left(\gamma_{\rm{th}}-1\right) \frac{N_o}{P_s}\right)^{-1}}-A_1\right) & \text{for }  \mathcal{C}_2 \\
                            1, & \text{for }\mathcal{C}_3
                        \end{array}
                \right.
                \label{Eq:P_o_final}
    \end{align} 
    \hrulefill
    \end{figure*}
    In~\eqref{Eq:P_o_final}, the conditions $\mathcal{C}_1$,  $\mathcal{C}_2$, and $\mathcal{C}_3$ respectively stand for 
    \begin{align}
       &  \mathcal{C}_1:  \frac{w_d}{\sqrt{2}} \sqrt{\frac{C_0}{\left|C_2\right|}}
        \nonumber \\ & \times
        \sqrt{\ln\left(\frac{1}{C_0}\left(\erf\left(\frac{\sqrt{2} \alpha}{w_d}\right)\right)^2 + \frac{1}{C_0} \left(\gamma_{\rm{th}}-1\right) \frac{N_o}{P_s}\right)^{-1}} < A_1,
    \end{align}
    \begin{align}
        &\mathcal{C}_2: A_1 \leq \frac{w_d}{\sqrt{2}} \sqrt{\frac{C_0}{\left|C_2\right|}}
        \nonumber \\ & \times
        \sqrt{\ln\left(\frac{1}{C_0}\left(\erf\left(\frac{\sqrt{2} \alpha}{w_d}\right)\right)^2 + \frac{1}{C_0} \left(\gamma_{\rm{th}}-1\right) \frac{N_o}{P_s}\right)^{-1}} \leq A_2
    \end{align}
    and
    \begin{align}
    &\mathcal{C}_3: \frac{w_d}{\sqrt{2}} \sqrt{\frac{C_0}{\left|C_2\right|}}
    \nonumber \\ & \times \sqrt{\ln\left(\frac{1}{C_0}\left(\erf\left(\frac{\sqrt{2} \alpha}{w_d}\right)\right)^2 + \frac{1}{C_0} \left(\gamma_{\rm{th}}-1\right) \frac{N_o}{P_s}\right)^{-1}} > A_2
    \end{align}
\end{prop}
\begin{IEEEproof}
    The outage probability is defined~as
    \begin{align}
        P_o = \Pr\left( C \leq r_{\rm{th}}\right),
        \label{Eq:P_o_definition}
    \end{align}
    where $C$ stands for the link capacity and can be obtained~as
    \begin{align}
       C = \log_2\left(\frac{h_b\,P_s}{N_o}+1\right),  
    \end{align}
    or, after applying~\eqref{Eq:h_b_2},
    \begin{align}
       C \approx \log_2\left(\frac{h_b^{(2)}\,P_s}{N_o}+1\right). 
       \label{Eq:C}
    \end{align}

From~\eqref{Eq:C},~\eqref{Eq:P_o_definition} can be rewritten~as
\begin{align}
        P_o = \Pr\left( \log_2\left(\frac{h_b^{(2)}\,P_s}{N_o}+1\right) \leq r_{\rm{th}}\right),
        \label{Eq:P_o_s1}
\end{align}
or
\begin{align}
        P_o = \Pr\left(h_b^{(2)} \leq \left(\gamma_{\rm{th}}-1\right) \frac{N_o}{P_s}\right),
\label{Eq:P_o_s1}
\end{align}
where 
\begin{align}
    \gamma_{\rm{th}} = 2r_{{\rm{th}}}. 
\end{align}
Next, we apply~\eqref{Eq:h_b_2} in~\eqref{Eq:P_o_s1} and we obtain
\begin{align}
    P_o & = \Pr\left( \left(\erf\left(\frac{\sqrt{2} \alpha}{w_d}\right)\right)^2 -  C_0 \exp\left(-\frac{2}{w_d^2} \frac{\left|C_2\right|}{C_0} r^2 \right)
    \right. 
    \nonumber \\ &  
    \leq \left. \left(\gamma_{\rm{th}}-1\right) \frac{N_o}{P_s}\right)
\end{align}
or equivalently
\begin{align}
    P_o  & = \Pr\left(
    \exp\left(-\frac{2}{w_d^2} \frac{\left|C_2\right|}{C_0} r^2 \right) \right.
    \nonumber \\ & 
    \geq \left. \frac{1}{C_0}\left(\erf\left(\frac{\sqrt{2} \alpha}{w_d}\right)\right)^2 + \frac{1}{C_0} \left(\gamma_{\rm{th}}-1\right) \frac{N_o}{P_s}\right), 
\end{align}
or
\begin{align}
    P_o  & = \Pr\left(
    -\frac{2}{w_d^2} \frac{\left|C_2\right|}{C_0} r^2 \right.
    \nonumber \\ & 
    \geq \left. \ln\left(\frac{1}{C_0}\left(\erf\left(\frac{\sqrt{2} \alpha}{w_d}\right)\right)^2 + \frac{1}{C_0} \left(\gamma_{\rm{th}}-1\right) \frac{N_o}{P_s}\right)\right),
\end{align}
or as in~\eqref{Eq:Po_final}, given at the top of the next page.
\begin{figure*}
\begin{align}
    P_o  & = \Pr\left(
     r
    \leq  \frac{w_d}{\sqrt{2}} \sqrt{\frac{C_0}{\left|C_2\right|}}\sqrt{\ln\left(\frac{1}{C_0}\left(\erf\left(\frac{\sqrt{2} \alpha}{w_d}\right)\right)^2 + \frac{1}{C_0} \left(\gamma_{\rm{th}}-1\right) \frac{N_o}{P_s}\right)^{-1}}\right)
    \label{Eq:Po_final}
\end{align}
\hrulefill
\end{figure*}
From~\eqref{Eq:Po_final}, we obtain~\eqref{Eq:Po_Fr}, given at the top of the next page.
\begin{figure*}
\begin{align}
    P_o = F_{r}\left(\frac{w_d}{\sqrt{2}} \sqrt{\frac{C_0}{\left|C_2\right|}}\sqrt{\ln\left(\frac{1}{C_0}\left(\erf\left(\frac{\sqrt{2} \alpha}{w_d}\right)\right)^2 + \frac{1}{C_0} \left(\gamma_{\rm{th}}-1\right) \frac{N_o}{P_s}\right)^{-1}}\right).
    \label{Eq:Po_Fr}
\end{align}
\hrulefill
\end{figure*}
Finally, by applying~\eqref{Eq:Fr} in~\eqref{Eq:Po_Fr}, we obtain~\eqref{Eq:P_o_final}. This concludes the~proof. 
\end{IEEEproof}

\section{Results \& Discussions}

This section focuses on presenting numerical results, which are verified through simulations, as well as fruitful discussions. The main goal is to extract insights about the probability and severity of the impact of blockage.  

\begin{figure}
	\centering
	\scalebox{1.0}{
		\begin{tikzpicture}[x=0.55pt,y=0.55pt,yscale=1,xscale=1]
			\begin{axis}[
				xlabel={$r\,\rm{(cm)}$},
				ylabel={$h_b$},
				legend pos=north east,
				legend style={legend cell align=left},
				xmin = 0,
				xmax =20,
				ymin = 0,
				ymax= 1,
				ymajorgrids=true,
				xmajorgrids=true,
				grid style=dashed,
				]
				\addplot[color=blue]
				coordinates {
					(0, 0.976727)
					(0.1, 0.976727)
					(0.5, 0.976727)
					(1.0, 0.976727)
					(1.5, 0.976728)
					(2.0, 0.976728)
					(2.5, 0.976729)
					(3.0, 0.97673)
					(3.5, 0.976731)
					(4.0, 0.976732)
					(4.5, 0.976733)
					(5.0, 0.976734)
					(5.5, 0.976736)
					(5.8, 0.976737)
					(6.0, 0.976738)
					(6.5, 0.97674)
					(7.0, 0.976741)
					(7.5, 0.976744)
					(8.0, 0.976746)
					(8.5, 0.976748)
					(9.0, 0.976751)
					(9.5, 0.976754)
					(10.0, 0.976756)
					(11.0, 0.976763)
					(15.0, 0.976793)
					(20.0, 0.976844)
				};
				\addlegendentry{{$f=1\,\rm{GHz}$}}
				
				\addplot[color=green]
				coordinates {
					(0, 0.121552)
					(0.01, 0.121552)
					(0.05, 0.121552)
					(0.1, 0.121552)
					(0.5, 0.121552)
					(0.8, 0.121553)
					(1.0, 0.121553)
					(2.0, 0.121553)
					(5.0, 0.12156)
					(8.0, 0.121571)
					(9.0, 0.121576)
					(10.0, 0.121582)
					(15.0, 0.121618)
					(16.0, 0.121627)
					(16.01, 0.121627)
					(16.1, 1.0)
					(16.3, 1.0)
					(16.5, 1.0)
					(17.0, 1.0)
					(20.0, 1.0)
				};
				\addlegendentry{{$f=10\,\rm{GHz}$}}
				
				\addplot[color=red]
				coordinates {
					(0, 0.0151889)
					(0.01, 0.0151889)
					(0.1, 0.0151889)
					(0.2, 0.0151889)
					(0.5, 0.015189)
					(1.0, 0.0151892)
					(2.0, 0.0151901)
					(3.0, 0.0151916)
					(4.0, 0.0151936)
					(5.0, 0.0151962)
					(6.0, 0.0151995)
					(7.0, 0.0152033)
					(8.0, 0.0152077)
					(8.5, 0.0152101)
					(8.51, 0.0152101)
					(8.55, 1.0)
					(8.6, 1.0)
					(8.9, 1.0)
					(9.0, 1.0)
					(10.0, 1.0)
					(20.0, 1.0)
				};
				\addlegendentry{{$f=20\,\rm{GHz}$}}

				\addplot[color=purple]
				coordinates {
					(0, 0.0170112)
					(0.01, 0.0170112)
					(0.05, 0.0170112)
					(0.1, 0.0170112)
					(0.2, 0.0170112)
					(0.5, 0.0170111)
					(1.0, 0.0170109)
					(2.0, 0.01701)
					(3.0, 0.0170085)
					(4.0, 0.0170065)
					(4.05, 1.0) 
					(4.1, 1.0)
					(4.5, 1.0)
					(5.0, 1.0)
				};
				\addlegendentry{{$f=50\,\rm{GHz}$}}
				
				\addplot[color=black]
				coordinates {
					(0, 0.0217026)
					(0.01, 0.0217026)
					(0.05, 0.0217026)
					(0.1, 0.0217026)
					(0.2, 0.0217026)
					(0.4, 0.0217025)
					(0.6, 0.0217025)
					(0.8, 0.0217024)
					(1.0, 0.0217023)
					(1.2, 0.0217021)
					(1.4, 0.021702)
					(1.6, 0.0217018)
					(1.8, 0.0217016)
					(2.0, 0.0217014)
					(2.2, 0.0217011)
					(2.5, 0.0217007)
					(2.51,1.0)
					(2.55, 1.0)
					(2.6, 1.0)
					(3.0, 1.0)
					(4.0, 1.0)
					(5.0, 1.0)
				};
				\addlegendentry{{$f=100\,\rm{GHz}$}}
			\end{axis}
		\end{tikzpicture}
		}
	\caption{$h_b$ vs $r$ for different values of $f$.}
	\label{Fig:fig1}
\end{figure}
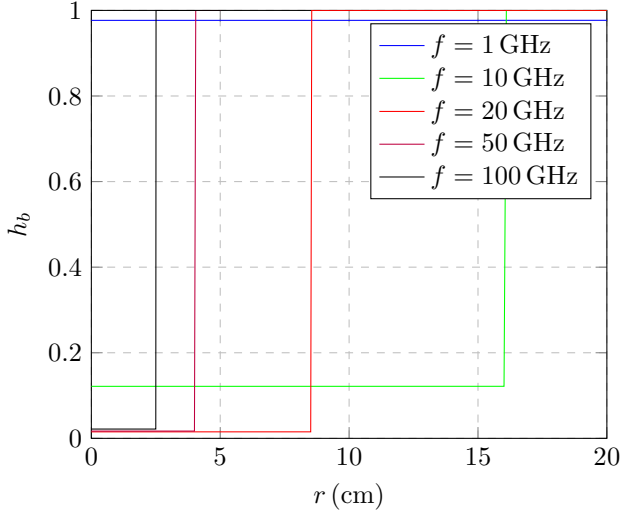
In Fig.~\ref{Fig:fig1}, $h_b$ is presented as a function of $r$, for different values of $f$, assuming that $\alpha_b=1\,\rm{cm}$, $G_r=30\,\rm{dBi}$, and transmission distance equal to $10\,\rm{m}$. Of note, in this result, $r$ is considered deterministic. As expected, for a given $f$, as $r$ increases, $h_b$ also increases. Moreover, we observe that as $f$ increases, the range of $r$ that affect $h_b$ decreases. For example, for $f=10\,\rm{GHz}$, $h_b$ is not equal to $1$ for $r$ in the range of $[0, 16.05\,\rm{cm}]$, while, for $f=100\,\rm{GHz}$, $h_b$ is not equal to $1$ for $r$ in the range of $[0, 2.5\,\rm{cm}]$. Finally, from this figure, it becomes evident that for the region in which $h_b$ is not equal to $1$, the impact of blockage increases as the operation frequency increase. In other words, it becomes apparent that as the frequency increases, the blockage probability decreases, since the transmission beam footprint at the receiver plane decreases, however, the impact of blockage become more severe.  

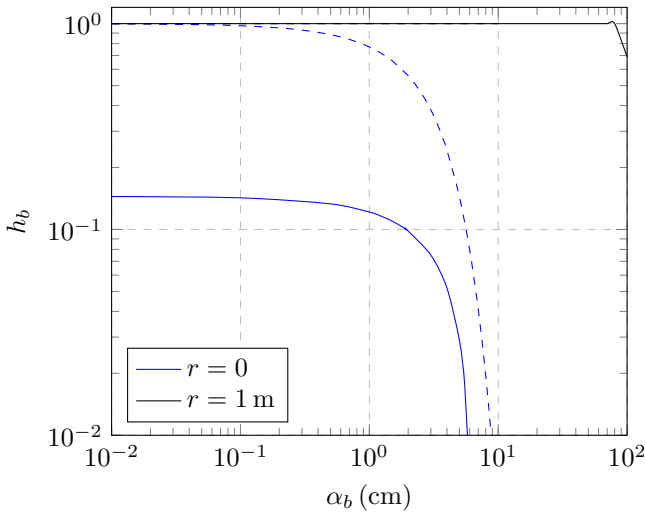
\begin{figure}
	\centering
	\scalebox{1.0}{
		
		\begin{tikzpicture}[x=0.55pt,y=0.55pt,yscale=1,xscale=1]
			\begin{axis}[
				xlabel={$\alpha_b\,\rm{(cm)}$},
				ylabel={$h_b$},
				legend pos=south west,
				legend style={legend cell align=left},
				xmin = 0.01,
				xmax =100,
				ymin = 0.01,
				ymax= 1.2,
				ymajorgrids=true,
				xmajorgrids=true,
				grid style=dashed,
				ymode=log,
				xmode=log,
				smooth
				]
				\addplot[
				color=blue]
				coordinates {
					(0.01, 0.144591)
					(0.1, 0.142496)
					(0.5, 0.133187)
					(1, 0.121552)
					(1.2, 0.1169)
					(1.5, 0.109923)
					(1.8, 0.102948)
					(1.9, 0.100624)
					(1.95, 0.0994624)
					(2, 0.0983006)
					(3, 0.0750884)
					(4, 0.0519354)
					(5, 0.0288611)
					(5.5, 0.0173594)
					(6.0, 0.0058847)
				};
				\addlegendentry{{$r=0$}}

				\addplot[
				color=black]
				coordinates {
					(0.01, 1)
					(0.1, 1)
					(0.5, 1)
					(1, 1)
					(10, 1)
					(20, 1)
					(30, 1)
					(40, 1)
					(50, 1)
					(60, 1)
					(70, 1)
					(80, 1)
					(100, 0.688313)
				};
				\addlegendentry{{$r=1\,\rm{m}$}}
				
				\addplot[
				color=blue,
				dashed]
				coordinates {
					(0.01, 0.99767)
					(0.02, 0.995339)
					(0.05, 0.988349)
					(0.1, 0.9767)
					(0.2, 0.95342)
					(0.3, 0.93018)
					(0.4, 0.906999)
					(0.5, 0.883897)
					(0.6, 0.860893)
					(0.7, 0.838007)
					(0.8, 0.815257)
					(0.9, 0.792662)
					(1.0, 0.77024)
					(1.1, 0.748008)
					(1.2, 0.725983)
					(1.4, 0.682623)
					(1.5, 0.661319)
					(2.0, 0.559138)
					(2.3, 0.501748)
					(2.5, 0.465296)
					(2.7, 0.430366)
					(2.9, 0.397007)
					(3.0, 0.38093)
					(3.2, 0.349996)
					(3.5, 0.306679)
					(3.8, 0.267071)
					(4.0, 0.242708)
					(5.0, 0.144206)
					(6.0, 0.0797104)
					(7.0, 0.040911)
					(8.0, 0.0194656)
					(9.0, 0.00857489)
					(10.0, 0.00349341)
				};
				\addplot[
				color=black,
				dashed]
				coordinates {
					(0.01, 0.999343)
					(0.02, 0.999629)
					(0.1, 1.0)
					(0.2, 1.0)
					(0.5, 1.0)
					(0.7, 1.0)
					(0.9, 1.0)
					(1.0, 1.0)
					(2.0, 1.0)
					(5.0, 1.0)
					(7.0, 1.0)
					(8.0, 1.0)
					(9.0, 1.0)
					(10.0, 1.0)
				};
			\end{axis}
		\end{tikzpicture}
		
		}
	\caption{$h_b$ vs $\alpha_b$ for different values of $r$ and $f=10\,\rm{GHz}$ (dashed lines) and $50\,\rm{GHz}$ (continuous lines).}
	\label{Fig:fig2}
\end{figure}

Figure~\ref{Fig:fig2} depicts $h_b$ as a function of $\alpha_b$ for different values of $r$ and $f$. As expected, for given $r$ and $f$, as $\alpha_b$ increases, the area of the shadow due to blockage increases; thus, $h_b$ decreases. Additionally, for fixed $f$ and $\alpha_b$, as $r$ increases,  the area of the shadow due to blockage decreases; hence, $h_b$ increases. Finally, for given $\alpha_b$ and $r$, as $f$ increases, the are of the transmission beam footprint at the receiver plane decreases; as a consequence, $h_b$ increases. 

\section{Conclusions}
In this contribution, we characterized the impact of blockage by presenting a partial blockage. We reported two low-complexity approximated expressions for the impact of blockage coefficient. To highlight the applicability of the approximations in complex environments, we documented the outage probability of a wireless link that suffer from partial blockage, for the case in which the distance between the center of the receiver plane and the blocker's shadow center follow uniform distribution. Numerical results verified our finding and revealed the impact of blockage under different setups and transmission parameters.  

\section*{Acknowledgment}
This work was supported by the MINOAS Project within the H.F.R.I call ``Basic Research Financing (Horizontal Support of all Sciences)'' through the National Recovery and Resilience Plan “Greece 2.0” funded by the European Union-NextGenerationEU (H.F.R.I.) under Project 15857.
\balance
\bibliographystyle{IEEEtran}
\bibliography{IEEEabrv,References}

\begin{thebibliography}{10}
\providecommand{\url}[1]{#1}
\csname url@samestyle\endcsname
\providecommand{\newblock}{\relax}
\providecommand{\bibinfo}[2]{#2}
\providecommand{\BIBentrySTDinterwordspacing}{\spaceskip=0pt\relax}
\providecommand{\BIBentryALTinterwordstretchfactor}{4}
\providecommand{\BIBentryALTinterwordspacing}{\spaceskip=\fontdimen2\font plus
\BIBentryALTinterwordstretchfactor\fontdimen3\font minus
  \fontdimen4\font\relax}
\providecommand{\BIBforeignlanguage}[2]{{%
\expandafter\ifx\csname l@#1\endcsname\relax
\typeout{** WARNING: IEEEtran.bst: No hyphenation pattern has been}%
\typeout{** loaded for the language `#1'. Using the pattern for}%
\typeout{** the default language instead.}%
\else
\language=\csname l@#1\endcsname
\fi
#2}}
\providecommand{\BIBdecl}{\relax}
\BIBdecl

\bibitem{10302317}
D.~J. Vergados, A.~Michalas, A.-A.~A. Boulogeorgos, S.~Nikolaou,
  N.~Asimopoulos, and D.~D. Vergados, ``Adaptive virtual reality streaming: A
  case for tcp,'' \emph{IEEE Transactions on Network and Service Management},
  vol.~21, no.~2, pp. 1518--1533, 2024.

\bibitem{9583918}
A.-A.~A. Boulogeorgos, J.~M. Jornet, and A.~Alexiou, ``Directional terahertz
  communication systems for 6g: Fact check,'' \emph{IEEE Vehicular Technology
  Magazine}, vol.~16, no.~4, pp. 68--77, 2021.

\bibitem{9615497}
T.~A. Tsiftsis, C.~Valagiannopoulos, H.~Liu, A.-A.~A. Boulogeorgos, and N.~I.
  Miridakis, ``Metasurface-coated devices: A new paradigm for energy-efficient
  and secure 6g communications,'' \emph{IEEE Vehicular Technology Magazine},
  vol.~17, no.~1, pp. 27--36, 2022.

\bibitem{9356523}
A.-A.~A. Boulogeorgos, S.~E. Trevlakis, and N.~D. Chatzidiamantis, ``Optical
  wireless communications for in-body and transdermal biomedical
  applications,'' \emph{IEEE Communications Magazine}, vol.~59, no.~1, pp.
  119--125, 2021.

\bibitem{Boulogeorgos2018}
A.-A.~A. Boulogeorgos, A.~Alexiou, T.~Merkle, C.~Schubert, R.~Elschner,
  A.~Katsiotis, P.~Stavrianos, D.~Kritharidis, P.~K. Chartsias, J.~Kokkoniemi,
  M.~Juntti, J.~Lehtom\"aki, A.~Teixeir\'a, and F.~Rodrigues, ``Terahertz
  technologies to deliver optical network quality of experience in wireless
  systems beyond {5G},'' \emph{IEEE Commun. Mag.}, vol.~56, no.~6, pp.
  144--151, Jun. 2018.

\bibitem{10528305}
A.-A.~A. Boulogeorgos, S.~E. Trevlakis, T.~A. Tsiftsis, and A.~Alexiou,
  ``Toward modeling and assessing the disorientation and misalignment effect in
  optical wireless nano-networks,'' \emph{IEEE Journal on Selected Areas in
  Communications}, vol.~42, no.~8, pp. 2009--2025, 2024.

\bibitem{Kizhakkekundil2021}
S.~Kizhakkekundil, J.~Morais, S.~Braam, and R.~Litjens, ``Four knife-edge
  diffraction with antenna gain model for generic blockage modelling,''
  \emph{IEEE Wireless Communications Letters}, vol.~10, no.~10, pp. 2106--2109,
  2021.

\bibitem{Alsaleem2020}
F.~Alsaleem, J.~S. Thompson, D.~I. Laurenson, S.~K. Podilchak, and C.~A.
  Alistarh, ``Small-size blockage measurements and modelling for mmwave
  communications systems,'' in \emph{2020 IEEE 31st Annual International
  Symposium on Personal, Indoor and Mobile Radio Communications}, 2020, pp.
  1--6.

\bibitem{Tang2020}
W.~Tang, J.~Y. Dai, M.~Z. Chen, K.-K. Wong, X.~Li, X.~Zhao, S.~Jin, Q.~Cheng,
  and T.~J. Cui, ``{MIMO} transmission through reconfigurable intelligent
  surface: {System} design, analysis, and implementation,'' \emph{IEEE J. Sel.
  Areas Commun.}, vol.~38, no.~11, pp. 2683--2699, Nov. 2020.

\bibitem{Gapeyenko2020}
M.~Gapeyenko, A.~Samuylov, M.~Gerasimenko, D.~Moltchanov, S.~Singh, M.~R.
  Akdeniz, E.~Aryafar, S.~Andreev, N.~Himayat, and Y.~Koucheryavy,
  ``Spatially-consistent human body blockage modeling: A state generation
  procedure,'' \emph{IEEE Transactions on Mobile Computing}, vol.~19, no.~9,
  pp. 2221--2233, 2020.

\bibitem{Wu2021}
Y.~Wu, J.~Kokkoniemi, C.~Han, and M.~Juntti, ``Interference and coverage
  analysis for terahertz networks with indoor blockage effects and
  line-of-sight access point association,'' \emph{IEEE Transactions on Wireless
  Communications}, vol.~20, no.~3, pp. 1472--1486, 2021.

\bibitem{Wang2023}
C.~Wang and Y.~J. Chun, ``Stochastic geometric analysis of the terahertz
  (thz)-mmwave hybrid network with spatial dependence,'' \emph{IEEE Access},
  vol.~11, pp. 25\,063--25\,076, 2023.

\bibitem{3GPP2024}
``Study on channel model for frequencies from 0.5 to 100 ghz(release 18),''
  3GPP, Tech. Rep., 2024.

\bibitem{B:Gra_Ryz_Book}
I.~S. Gradshteyn and I.~M. Ryzhik, \emph{Table of Integrals, Series, and
  Products}, 6th~ed.\hskip 1em plus 0.5em minus 0.4em\relax New York: Academic,
  2000.

\bibitem{A:Analytical_Performance_Assessment_of_THz_Wireless_Systems}
A.-A.~A. Boulogeorgos, E.~N. Papasotiriou, and A.~Alexiou, ``Analytical
  performance assessment of {THz} wireless systems,'' \emph{IEEE Access},
  vol.~7, no.~1, pp. 1--18, Jan. 2019.

\bibitem{Farid2007}
A.~A. Farid and S.~Hranilovic, ``Outage capacity optimization for free-space
  optical links with pointing errors,'' \emph{Journal of Lightwave Technology},
  vol.~25, no.~7, pp. 1702--1710, Jul. 2007.

\bibitem{papoulis}
A.~Papoulis and S.~Pillai, \emph{Probability, Random Variables, and Stochastic
  Processes}, ser. McGraw-Hill series in electrical engineering: Communications
  and signal processing.\hskip 1em plus 0.5em minus 0.4em\relax Tata
  McGraw-Hill, 2002.

\end{thebibliography}

\end{document}